%% file: main.tex
\documentclass[conference, 10pt]{IEEEtran}
\usepackage[utf8]{inputenc}
\usepackage{cite}
\usepackage{verbatim}
\usepackage{graphicx}
\usepackage{amsmath}
\usepackage{algorithm,algorithmic}
\usepackage{tikz}
\usepackage{url}
\usepackage{subcaption}
\usepackage[bottom]{footmisc}
\usetikzlibrary{shapes,arrows,matrix,positioning}
\usetikzlibrary{decorations.pathreplacing}
\usetikzlibrary{backgrounds}

\tikzstyle{block} = [draw, fill=blue!20, rectangle, 
    minimum height=3em, minimum width=6em]
\tikzstyle{sum} = [draw, fill=blue!20, circle, node distance=1cm]
\tikzstyle{var_node} = [draw, fill=red!20, circle, node distance=0.5cm]
\tikzstyle{check_node} = [draw, fill=blue!20, rectangle, node distance=0.5cm]
\tikzstyle{input} = [coordinate]
\tikzstyle{output} = [coordinate]
\tikzstyle{pinstyle} = [pin edge={to-,thin,black}]

\title{Distributed Processing for Encoding and Decoding of Binary LDPC codes using MPI}
\author{\IEEEauthorblockN{Bhargav Gokalgandhi\IEEEauthorrefmark{1}, Ivan Seskar\IEEEauthorrefmark{2} \\}
\IEEEauthorblockA{WINLAB, Rutgers University\\
671, US-1, North Brunswick Township, NJ 08902 \\
Email: \IEEEauthorrefmark{1}bgokal@winlab.rutgers.edu, \IEEEauthorrefmark{2}seskar@winlab.rutgers.edu}
\thanks{This project was funded by the NSF "COSMOS" Project under grant number CNS-1827923.}
}

\IEEEoverridecommandlockouts
\begin{document}

\IEEEpubid{\makebox[\columnwidth]{978-1-7281-1878-9/19/\$31.00~\copyright2019 IEEE \hfill} \hspace{\columnsep}\makebox[\columnwidth]{ }}

\maketitle

\IEEEpubidadjcol

\begin{abstract}
    Low Density Parity Check (LDPC) codes are linear error correcting codes used in communication systems for Forward Error Correction (FEC). But, intensive computation is required for encoding and decoding of LDPC codes, making it difficult for practical usage in general purpose software based signal processing systems. In order to accelerate the encoding and decoding of LDPC codes, distributed processing over multiple multi-core CPUs using Message Passing Interface (MPI) is performed. Implementation is done using Stream Processing and Batch Processing mechanisms and the execution time for both implementations is compared w.r.t variation in number of CPUs and number of cores per CPU. Performance evaluation of distributed processing is shown by variation in execution time w.r.t. increase in number of processors (CPU cores).
\end{abstract}

\input{Introduction.tex}

\input{Algorithms.tex}
\input{Implementation_Details.tex}
\input{Performance_Evaluation.tex}
\input{Conclusion.tex}

%\nocite{*}
\bibliographystyle{ieeetr}
%\printbibliography
\bibliography{ref.bib}

\end{document}

%% file: Introduction.tex
\section{Introduction}\label{sec:intro}

Low Density Parity Check (LDPC) codes are a type of Error Correcting codes which were developed by Robert Gallager in early 1960s \cite{gallager_ldpc}. Currently, LDPC codes are used widely for Forward Error Correction (FEC) in current and next-gen wireless standards such as 5G, 802.11, etc, as well as in video broadcasting, due to their capacity approaching performance with increase in block length. But, with increasing block length, the computation time of encoding and decoding processes of LDPC codes increases leading to high latency and limitation in throughput. 
%Also, M-QAM (M-ary Quadrature Amplitude Modulation) is a widely used channel modulation scheme to get higher throughput in communication systems. when M-QAM modulation is used in combination with Error Correcting Codes, it adds processing latency for channel encoding and decoding in practical scenarios. 
Considering the increasing usage of LDPC codes, and the current focus in 'softwarization' of front-end signal processing and networking elements for wireless systems, accelerating and distributing the computation for encoding and decoding of LDPC codes becomes important to achieve practical usage in next-gen software defined communication systems.

%In practice, the computational complexity of encoding and decoding process of LDPC codes varies based on different aspects such as rate, block length, matrix density, etc.
%The real-time implementation of LDPC codes in practice is done by ASICs (Application Specific Integrated Circuits) and hardware architecture designed specifically for processing of LDPC codes. 
Over the years, various methods have been used for accelerating the processing of LDPC codes using general purpose hardware such as multi-core Central Procssing Units (CPUs) and General Purpose Graphics Processing Units (GPGPUs). \cite{zhen2015parallel, zhao2014implementation, lin2014high, jiang2012efficient, kang2012parallel, chang2011accelerating, falcao2011massively, ji2009massively, wang2008parallel} use GPGPU, or a combination of GPGPU and OpenMP for acceleration of encoding and Sum-Product decoding of Binary LDPC codes. These papers also compare the GPGPU and multi-core CPU implementations in terms of acceleration provided. But the above papers show implementation of LDPC encoding and decoding on single CPU with shared memory parallelization model, and single GPGPU. While such systems provide good acceleration, it becomes difficult to distribute the processing to multi-server and multi-CPU systems.
%Also, while GPUs can be used for massive parallelization. the difficulty of accelerating iterative processes, and the memory limit of GPUs, limits the performance of using a single GPU for processing.
In such cases distributed memory mechanisms with message passing models can be used for ease of processing distribution over multiple CPUs for providing acceleration. LDPC encoding and decoding using message passing model has not been implemented or evaluated as much. In \cite{nittoor2011parallelizing} distributed LDPC decoding is done using MPI, but it aims at optimization of coarse grain graph search problem to select the LDPC Parity Check matrix with least Bit Error Rate (BER). It does not evaluate the computational performance of encoding and decoding LDPC codes.  %Furthermore, the above papers do not consider M-QAM modulation which create additional latency for encoding and soft-decision decoding required for combination of LDPC and M-QAM based processing.

In this paper, we use distributed memory message passing model by using Message Passing Interface (MPI) to distribute the encoding and Sum-Product decoding processes of LDPC codes over multiple multi-core CPUs. We perform the processing using Stream Processing and Batch Processing mechanisms. We show the acceleration provided by using message passing model for both mechanisms, and compare the execution time required for processing w.r.t. increase in number of CPUs and cores per CPU.% We show the advantages and limitations of both methods for encoding and decoding of LDPC codes.

The paper is organized as follows. Section \ref{sec:algos} describes the theoretical background related to encoding and decoding of LDPC codes. Section \ref{sec:impl} describes the various components used for experimental evaluation, and it explains the distributed implementation for each part of the algorithms mentioned in Section \ref{sec:algos}. Section \ref{sec:eval} shows the results of the various experiments conducted. Lastly, Section \ref{sec:conc} concludes the paper and states future directions to be taken.

%% file: Algorithms.tex
\section{Theoretical Background}\label{sec:algos}

%LDPC codes are a type of linear block codes which are used for error correction by adding redundancy to information bits that are to be encoded. 
We consider a vector $\mathrm{m}$ of $k$ information bits $[m_1, m_2, .... m_k]$. Using LDPC codes, $k$ information bits can be encoded into a vector $\mathrm{p}$ of $n$ bits $[p_1, p_2, ..., p_n]$, with $n \geq k$. The $n$ bits consist of $k$ information bits from vector $m$ and $n - k$ parity bits. So the rate of code is defined as $Rate(R) \geq k/n$. Here, $n > k > 0$.

The parity bits are formed by a linear combination of the $k$ information bits known as parity check equations. LDPC codes are defined by putting these parity check equations in a Parity Check matrix, denoted by $\mathrm{H}$, which is a $k \times (n-k)$ matrix. Each row of $\mathrm{H}$ denotes a parity check equation $c$, and each column defines whether a bit from the vector $\mathrm{p}$ is present in the parity check equation or not. Depending on the number of ones in each row and column, the Parity Check matrix is defined as $(n_c, n_v)$, where $n_c$ is the number of ones per column, and $n_v$ is the number of ones per row. Then, the rate of code can be defined as $Rate(R) \geq 1 - n_c/n_v$. LDPC codes with this type of matrix are known as regular LDPC codes.

\begin{equation}\label{eq:hp_0}
    \mathrm{H}\mathrm{p}^T = 0
\end{equation}

Where $T$ denotes transpose of a vector or a matrix. This means that the additional parity bits must take values $0$ or $1$ such that each parity check equation equals $0$.

\subsection{LDPC Encoding}\label{subsec:ldpc_encode}

From Eq. \ref{eq:hp_0}, we can say that each codeword is present in the null space of the Parity Check matrix. So, we form the Generator matrix $\mathrm{G}$, which is a $k \times n$ matrix, for encoding of LDPC codes by using the equation,

\begin{equation}\label{eq:HG}
    \mathrm{HG}^T = 0
\end{equation}

A Generator matrix is created by using Eq. \ref{eq:HG} i.e. by finding the null space of the Parity Check matrix. Then, by using the Generator matrix encoding of the input information bits can be done using,

\begin{equation}\label{eq:encoding}
    \mathrm{m}^T\mathrm{G} = \mathrm{p}^T
\end{equation}

\subsection{LDPC Decoding}\label{subsec:ldpc_decode}

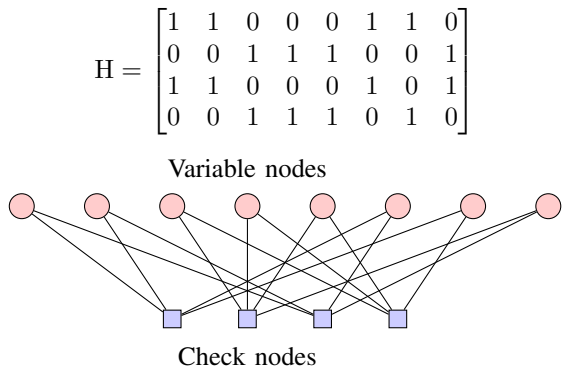
\begin{figure}[t]
    \begin{center}
    $\mathrm{H} = 
        \begin{bmatrix}
        1&1&0&0&0&1&1&0\\
        0&0&1&1&1&0&0&1\\
        1&1&0&0&0&1&0&1\\
        0&0&1&1&1&0&1&0\\
        \end{bmatrix}$
    \end{center}
    \begin{center}
        \begin{tikzpicture}[auto, node distance=2cm,>=latex']
            % We start by placing the blocks
            %\node [sum, right of=input] (sum) {};
            \node (in_text) at(3, 0.5) {Variable nodes};
            \node [var_node] at(0,0) (v1) {};
            \node [var_node] at(1,0) (v2) {};
            \node [var_node] at(2,0) (v3) {};
            \node [var_node] at(3,0) (v4) {};
            \node [var_node] at(4,0) (v5) {};
            \node [var_node] at(5,0) (v6) {};
            \node [var_node] at(6,0) (v7) {};
            \node [var_node] at(7,0) (v8) {};
            \node [check_node] at(2,-1.5) (c1) {};
            \node [check_node] at(3,-1.5) (c2) {};
            \node [check_node] at(4,-1.5) (c3) {};
            \node [check_node] at(5,-1.5) (c4) {};
            \node (in_text) at(3, -2) {Check nodes};
            % We draw an edge between the controller and system block to 
            % calculate the coordinate u. We need it to place the measurement block. 
            %\node [output, right of=encoder] (output) {};
           % \node [right of=output] (out_text) at (3,0.25) {$p_1, p_2, p_3, ..., p_n$};
        
            % Once the nodes are placed, connecting them is easy. 
            %\draw [draw,->] (input) -- node {} (encoder);
            %\draw [->] (encoder) -- node {} (output);
            \draw (v1) -- node {} (c1);
            \draw (v1) -- node {} (c3);
            \draw (v2) -- node {} (c3);
            \draw (v2) -- node {} (c1);
            \draw (v3) -- node {} (c2);
            \draw (v3) -- node {} (c4);
            \draw (v4) -- node {} (c2);
            \draw (v4) -- node {} (c4);
            \draw (v5) -- node {} (c2);
            \draw (v5) -- node {} (c4);
            \draw (v6) -- node {} (c1);
            \draw (v6) -- node {} (c3);
            \draw (v7) -- node {} (c1);
            \draw (v7) -- node {} (c4);
            \draw (v8) -- node {} (c2);
            \draw (v8) -- node {} (c3);
        \end{tikzpicture}
    \end{center}
    \caption{Bipartite graph for decoding of LDPC codes for a random $(2,4)$ Parity Check matrix with $n=8$ and $Rate(R) \geq 1/2$}\label{fig:graph}
\end{figure}

We have taken the steps of Sum-Product decoding algorithm from \cite{gallager_ldpc, johnson2006introducing, johnson2002low, shokrollahi2003ldpc}. The aforementioned papers contain the sum-product algorithm description in detail. We explain the gist of the algorithm. For decoding, a bipartite graph $\mathcal{G(V, C, E)}$ is created from the Parity Check matrix, where $\mathcal{V}$ is a set of $n$ variable nodes with index $v$, $\mathcal{C}$ is a set of $n - k$ check nodes with index $c$. If there is a $1$ on the $i^{th}$ row and $j^{th}$ column of the Parity Check matrix, an edge connects the $i^{th}$ check node and $j^{th}$ variable node. An example is shown in Fig. \ref{fig:graph}. We represent the variable nodes using $v$ and check nodes using $c$. The variable nodes and check nodes transfer messages and update Log-Likelihood Ratio (LLR) values of each bit.
%At first, the LLR values are taken using QAM demapper. 
These LLR values are input to the variable nodes as,

\begin{equation}\label{eq:in_llr}
    L_v = R_v = \frac{\text{prob(bit}_v = 1)}{\text{prob(bit}_v = 0)} = \frac{2y_i}{\sigma^2}
\end{equation}

where $L_v$ is the total LLR value and $R_v$ is the input LLR value of node $v \in \mathcal{V}$, prob(.) is the probability value, $y_i$ is the received noisy bit value, $\sigma^2$ is the Additive White Gaussian Noise (AWGN) power at the receiver. 
%More explanation regarding how the LLR values are obtained will be shown in subsection \ref{subsec:qam_map}.
These LLR values are sent to all the check nodes connected to each variable nodes. After the check nodes receive the LLR values, the check node values are updated using the equation,

\begin{equation}\label{eq:extrin}
    E_{cv} = log_e\left(\frac{1 + \prod_{\hat{v} \in \mathcal{V}, \hat{v} \neq v}tanh(L_{\hat{v}c}/2)}{1 + \prod_{\hat{v} \in \mathcal{V}, \hat{v} \neq v}tanh(L_{\hat{v}c}/2)}\right)
\end{equation}

where $E_{cv}$ is the partial LLR update that each check node $c \in \mathcal{C}$ calculates for variable node $v$, and $L_{\hat{v}c}$ is the partial LLR update each variable node $v$ calculates for check node $c$ using the value from Eq. \ref{eq:extrin},

\begin{equation}\label{eq:intrin}
    L_{vc} = \sum_{\hat{c} \in \mathcal{C}, \hat{c} \neq c} E_{\hat{c}v} + R_v
\end{equation}

After a predefined number of iterations, using the values gained from Eq. \ref{eq:extrin}, the total LLR value is calculated using equation,

\begin{equation}\label{ref:total_llr}
    L_{v} = \sum_{c \in C} E_{cv} + R_v
\end{equation}

and then a hard-decision for each bit is made using,

\begin{equation}\label{eq:hard_decision}
    z_v = \left\{
      \begin{array}{lr}
        1 & : L_v \geq 0\\
        0 & : L_v < 0
      \end{array}
    \right.
\end{equation}

where $z_v$ is the final bit value for variable node $v$.

%% file: Implementation_Details.tex
\section{Implementation Details}\label{sec:impl}

\subsection{Components used for implementation}\label{subsec:comps_impl}

\subsubsection{{ORBIT Testbed}}

Open-Access Research Testbed (ORBIT) is a large-scale academic indoor wireless testbed consisting of a 20x20 grid of computing nodes, more than 100 Software Defined Radios (SDRs), and a set of inter-connected servers for large-scale distributed computing and high performance computing applications. All components are connected to a centralized server system using which control functions and data transfer can be performed. More information related to the testbed can be found in \cite{raychaudhuri2005overview, orbiturl}.

\subsubsection{{Intel(R) Xeon(R) CPU E5-2698 v3 @ 2.30GHz}}

This CPU has 16 physical cores and 32 logical cores. It has a base frequency of 2.3 GHz and max turbo frequency of 3.6 GHz, maximum memory bandwidth of 68 GBps with a 40 MB cache for fast memory access. More information can be seen at \cite{intelurl}. We use 2 server nodes, each consisting of 2 CPUs. These server nodes are present in ORBIT testbed. Both servers are connected using a 25 Gigabit ethernet link which will be utilized when multi-server distributed processing is implemented.

\subsubsection{MPI}

MPI is a standardized interface for message passing between processors in applications running on distributed memory systems. MPI can be used by processes running independently on clusters of multiple CPUs for communicating with each other and passing data using messages. We use MPICH \cite{mpichurl}, an open-source and widely portable implementation of MPI standard for distributing the processing among multiple cores and multiple servers. Some communication routines which are used in this paper for implementation are,
\begin{itemize}
    \item \textit{MPI\_Send(): } Used to send data from a specific process to another.
    \item \textit{MPI\_Recv(): } Used to receive data from a specific process.
    \item \textit{MPI\_Gatherv(): } Used to gather data from all processes to one. Each process can contain variable amount of data to be gathered.
    \item \textit{MPI\_Bcast(): } Used to broadcast data from one process to all other processes.
\end{itemize}

\subsection{Distributed Stream Processing implementation}\label{subsec:stream_proc}

In Stream Processing, a single vector from a stream of multiple vectors is computed in a distributed manner. The data within each vector is distributed among multiple processors which perform computations and then the result is gathered in the master processor. We assume the availability of $N_{proc}$ processors denoted by $Proc(0), Proc(1), ..., Proc(N_{proc} - 1)$, where $Proc(x)$ means processor with index $x$. The division of elements to be processed among the processors is done in the following manner,

\begin{equation}\label{eq:elems_div}
    NumElems(q) = \left\{
      \begin{array}{ll}
        floor(N_{proc}/l) + 1 & : x < N_{procs} \% l \\
        floor(N_{proc}/l) & : x \geq N_{procs} \% l
      \end{array}
    \right.
\end{equation}

where $NumElems(q)$ represent the number of elements to be given to processor $q$, $a\%b$ means the remainder of $a/b$, and $l$ is the total number of elements to be divided.

\subsubsection{{Encoding}}

\begin{figure}[t]
    \centering
    \begin{tikzpicture}
        \node at(0, 0) {
        $\mathrm{m}^T\mathrm{G} = 
        \begin{bmatrix}
            1&0&1&1
        \end{bmatrix}
        \begin{bmatrix}
        1&0&0&0&\textcolor{red}0&\textcolor{blue}1&\textcolor{olive}0&\textcolor{purple}0\\
        0&1&0&0&\textcolor{red}1&\textcolor{blue}0&\textcolor{olive}0&\textcolor{purple}0\\
        0&0&1&0&\textcolor{red}0&\textcolor{blue}0&\textcolor{olive}1&\textcolor{purple}0\\
        0&0&0&1&\textcolor{red}1&\textcolor{blue}1&\textcolor{olive}1&\textcolor{purple}1\\
        \end{bmatrix}
         = \mathrm{p}^T$
        };
        \node[color=red] (proc0) at(-3, -2) {$Proc(0)$};
        \node[color=blue] (proc1) at(-1, -2) {$Proc(1)$};
        \node[color=olive] (proc2) at(1, -2) {$Proc(2)$};
        \node[color=purple] (proc3) at(3, -2) {$Proc(3)$};
        %\draw [decorate,decoration={brace,amplitude=5pt,mirror,raise=4ex}] (0.45,-0.1) -- (0.9, -0.1) node [midway,yshift=-3em];
        %\draw [decorate,decoration={brace,amplitude=5pt,mirror,raise=4ex}] (0.35.-0.1) --++ (0.4, 0) node [midway,yshift=-3em];
        \draw[->,dashed,color=red] (proc0.north) -- (1.5, -0.8) node {};
        \draw[->,dashed,color=blue] (proc1.north) -- (2, -0.8) node {};
        \draw[->,dashed,color=olive] (proc2.north) -- (2.6, -0.8) node {};
        \draw[->,dashed,color=purple] (proc3.north) -- (3.1, -0.8) node {};
    \end{tikzpicture}
    
    \begin{tikzpicture}
    \node at(-2.5, -1.5) (mult0) {
        $\textcolor{red}{p_4} = 
        \begin{bmatrix}
            1&0&1&1
        \end{bmatrix}
        \begin{bmatrix}
            \textcolor{red}0\\
            \textcolor{red}1\\
            \textcolor{red}0\\
            \textcolor{red}1\\
        \end{bmatrix}$
    };
    \node [below of = mult0, node distance = 1 cm, color=red] {$Proc(0)$};
    
    \node [right of = mult0, node distance = 5 cm] (mult1) {
        $\textcolor{blue}{p_5} = 
        \begin{bmatrix}
            1&0&1&1
        \end{bmatrix}
        \begin{bmatrix}
            \textcolor{blue}1\\
            \textcolor{blue}0\\
            \textcolor{blue}0\\
            \textcolor{blue}1\\
        \end{bmatrix}$
    };
    \node [below of = mult1, node distance = 1 cm, color=blue] {$Proc(1)$};
    
    \node [below of = mult0, node distance = 3 cm] (mult2) {
        $\textcolor{olive}{p_6} = 
        \begin{bmatrix}
            1&0&1&1
        \end{bmatrix}
        \begin{bmatrix}
            \textcolor{olive}0\\
            \textcolor{olive}0\\
            \textcolor{olive}1\\
            \textcolor{olive}1\\
        \end{bmatrix}$
    };
    \node [below of = mult2, node distance = 1 cm, color=olive] {$Proc(2)$};
    
    \node [right of = mult2, node distance = 5 cm] (mult3) {
        $\textcolor{purple}{p_7} = 
        \begin{bmatrix}
            1&0&1&1
        \end{bmatrix}
        \begin{bmatrix}
            \textcolor{purple}0\\
            \textcolor{purple}0\\
            \textcolor{purple}0\\
            \textcolor{purple}1\\
        \end{bmatrix}$
    };
    \node [below of = mult3, node distance = 1 cm, color=purple] {$Proc(3)$};
    
   %\draw[->, dashed, color=blue] (mult0.west) edge [bend right=75] (3,-5) node {} (final_mult);
    
    \end{tikzpicture}
    
    \begin{comment}
    \hfill
    \begin{tikzpicture}
    \node (non_mult) at(0, 0) {
        \begin{bmatrix}
            $m_0$&$m_1$&$m_2$&$m_3$
        \end{bmatrix}
        $ = $
        \begin{bmatrix}
            $p_0$&$p_1$&$p_2$&$p_3$
        \end{bmatrix}
        $ = $
        \begin{bmatrix}
            1&0&1&1
        \end{bmatrix}
    };
    \end{tikzpicture}
    \end{comment}
    
    \hfill
    
    $\mathrm{p}^T
     = 
    \begin{bmatrix}
        p_0&p_1&p_2&p_3&\textcolor{red}{p_4}&\textcolor{blue}{p_5}&\textcolor{olive}{p_6}&\textcolor{purple}{p_7}
    \end{bmatrix}$ \\
    \vspace{0.25cm}
    $ = 
    \begin{bmatrix}
        1&0&1&1&\textcolor{red}1&\textcolor{blue}0&\textcolor{olive}0&\textcolor{purple}1
    \end{bmatrix}$
    
    \caption{Example of Distributed Matrix-Vector multiplication for LDPC Encoding with a random Generator matrix.}
    \label{fig:encoding_proc}
\end{figure}
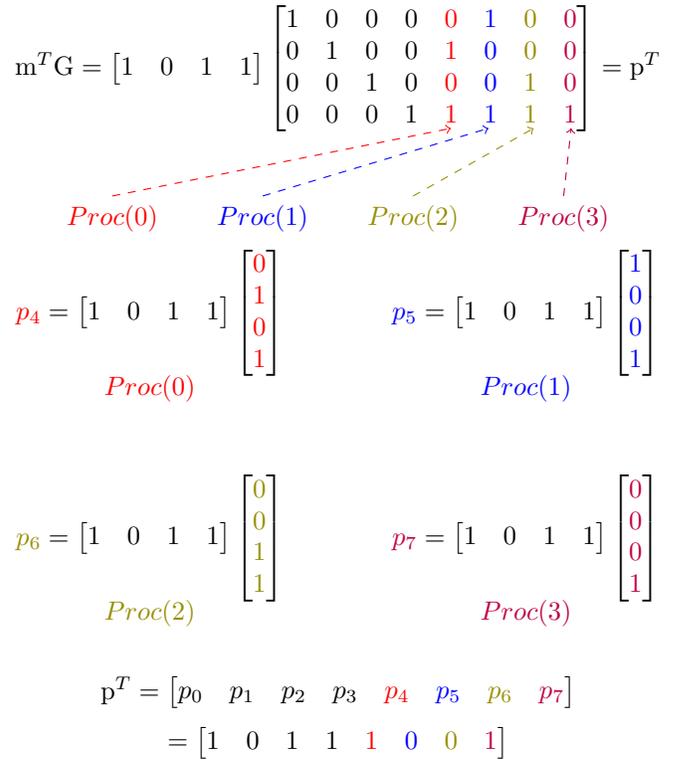

For encoding, we use the Generator matrix which is converted to standard $[I_k:P]$ where $I_k$ sub-matrix is a $k \times k$ identity matrix formed by column permutations of Generator matrix, and $P$ is a sub-matrix consisting of remaining $k \times (n-k)$ values of the Generator matrix. The input information bits $m_1,...,m_k$ are first copied to the output vector bits $p_1,..,p_k$ for each processor. Then, the columns of the $P$ sub-matrix of the Generator matrix are divided among the $N_{proc}$ processors using Eq. \ref{eq:elems_div}. An example of division of $4$ columns of the $P$ matrix is shown in Fig. \ref{fig:encoding_proc}. Each processor selects a subset of columns from the $k$ columns of sub-matrix $P$. Each processor then takes the input vector $\mathrm{m}$ and multiplies it with the columns of sub-matrix $P$ selected by that processor. The partial output calculated by each processor is then gathered at the master processor by using the \textit{MPI\_Gatherv()} command for further processing. The encoded bits are then converted to Binary Phase Shift Keying (BPSK) form for transmission through AWGN channel. For BPSK, we convert the $0$ bits to $-1$ and $1$ bits remain as $1$. %The final output vector $\mathrm{p}$ then consists of $[\mathrm{m}^T:\mathrm{m}^T P]$ values which are mapped using M-QAM mapper.

\begin{comment}
\subsubsection{{M-QAM Mapping and Demapping}}

The encoded bits are divided among the processors using Eq. \ref{eq:elems_div}, with the total number of elements $l = n/log_2(M)$. Then, each processor maps the partial vector using the algorithm in Section \ref{subsec:qam_map}. The partially mapped output is then gahtered from each process to the master process by using \textit{MPI\_Gatherv()}. After mapping and gather operations are done, the LDPC encoded vector $\mathrm{p}$ of length $n$ gets converted to a mapped vector of length $ceil(n/log_2(M))$, denoted by $\mathrm{s}$. The partial output vector is then  which is transmitted through AWGN channel. For demapping, the noisy symbols received at the M-QAM demapper, denoted by $\mathrm{y}$, are divided using Eq. \ref{eq:elems_div}, where $l = n/log_2(M)$. Each processor takes the partial noisy vector as input and calculates LLR values using Eq. \ref{eq:qam_to_llr}. Then, the partially LLR vector $R_v$ is gathered at the master process using \textit{MPI\_Gatherv()} and broadcast to all processes \textit{MPI\_Bcast()}. The input LLR values are then used for Sum-Product Decoding of LDPC codes.
\end{comment}

%\bigskip

%The result is then transmitted through an emulated AWGN channel i.e. Gaussian Noise is added to each bit.

%\bigskip

\subsubsection{{Decoding}}

\begin{algorithm}[t]
    \caption{Distributed Sum-Product Decoding using MPI}\label{alg:decoding}
    \begin{algorithmic}[1]
        \STATE {Define $\mathcal{V}$ and $\mathcal{C}$ as sets containing variable nodes $v$ and check nodes $c$ respectively, $E_{cv}, L_{vc}, L_v, R_v, z_v$,  $N_{proc}$, $Proc(x)$}
        %\STATE {Start interleaved execution:}
        \STATE {Divide the check and variable nodes to all processors in an interleaved manner}
            \FOR {$v \in \mathcal{V}$}
                \STATE {Find the initial LLR values $L_v$ and $R_v$ and broadcast to all processors using  \textit{MPI\_Bcast()}}
            \ENDFOR
            \FOR {$i = 1$ to \#iterations}
                \STATE {// Updating LLR values at check nodes}
                \FOR {($c = 0; c < |\mathcal{C}|; c = c + N_{proc}$)}
                    \STATE {Find $E_{(c+x)v}$}
                \ENDFOR
                \FOR {($c = 0; c < |\mathcal{C}|; c = c + 1$)}
                    \STATE {Send to processor which has the variable node $v$ in its list, (i.e. $Proc(v\%N_{proc})$, using \textit{MPI\_Send()}})
                    \STATE {Receive the message using \textit{MPI\_Recv()} at $Proc(v\%N_{proc})$}
                \ENDFOR
                
                \STATE {// Updating LLR values at variable nodes}
                \FOR {($v = 0; v < $size of V$; v = v + N_{proc}$)}
                    \STATE {Find $L_{(v+x)c}$}
                \ENDFOR
                \FOR {($v = 0; v < |\mathcal{V}|; v = v + 1$)}
                    \STATE {Send to processor which has the check node $c$ in its list, (i.e. $Proc(c\%N_{proc})$, using \textit{MPI\_Send()}})
                    \STATE {Receive the message using \textit{MPI\_Recv()} at $Proc(c\%N_{proc})$}
                \ENDFOR
            \ENDFOR
            
            \STATE {// Calculating final LLR values and then performing hard-decision decoding}
            \FOR {($v = 0; v < |\mathcal{V}|; v = v + N_{proc}$)}
                \STATE {Calculate $L_{v+x}$}
                \STATE {Broadcast output to all processes using \textit{MPI\_Bcast()}}
                \STATE {Calculate $z_{v+x}$ value}
                \STATE {Gather at master processor using \textit{MPI\_Gatherv()}}
            \ENDFOR
        \end{algorithmic}
\end{algorithm}

\begin{figure}[t]
    \begin{center}

        \begin{tikzpicture}[auto, node distance=2cm,>=latex']
            % We start by placing the blocks
            %\node [sum, right of=input] (sum) {};
            %\node (in_text) at(3, 0.5) {Variable nodes};
            \node [var_node] at(0,0) (v1) {};
            \node[above of = v1, node distance = 0.2 cm, anchor = south west] (in_text) at(0,0.5) (l1) {$Proc(0)$:\textit{MPI\_Send()}};
            \node [var_node] at(1,0) (v2) {};
            \node [var_node] at(2,0) (v3) {};
            \node [var_node] at(3,0) (v4) {};
            \node [var_node] at(4,0) (v5) {};
            \node [var_node] at(5,0) (v6) {};
            \node [var_node] at(6,0) (v7) {};
            \node [var_node] at(7,0) (v8) {};
            \node [check_node] at(2,-1.5) (c1) {};
            \node [check_node] at(3,-1.5) (c2) {};
            \node [check_node] at(4,-1.5) (c3) {};
            \node [check_node] at(5,-1.5) (c4) {};
            %\node(in_text) at(2,-2) (l2) {$L_0$};
            \node[below of = c3, node distance = 0.33 cm] (out_text) at(3,-2) (l3) {$Proc(2)$:\textit{MPI\_Recv()}};
            \node at(3, -3) (sub_cap) {Variable node partial LLR ($L_{vc}$) update and data transfer};
            %\node(in_text) at(4,-2) (l4) {$L_0$};
            %\node (in_text) at(3, -2) {Check nodes};
            % We draw an edge between the controller and system block to 
            % calculate the coordinate u. We need it to place the measurement block. 
            %\node [output, right of=encoder] (output) {};
           % \node [right of=output] (out_text) at (3,0.25) {$p_1, p_2, p_3, ..., p_n$};
        
            % Once the nodes are placed, connecting them is easy. 
            %\draw [draw,->] (input) -- node {} (encoder);
            %\draw [->] (encoder) -- node {} (output);
            \draw[->] (l1) -- node {} (v1);
            %\draw[->] (v1) -- node {} (c1);
            \draw[->] (v1) -- node {$L_{02}$} (c3);
            \draw[->] (c3) -- node {} (l3);
            %\draw[->] (v1) -- node {} (c3);
        \end{tikzpicture}
        
        \hfill
        
        \begin{tikzpicture}[auto, node distance=2cm,>=latex']
            % We start by placing the blocks
            %\node [sum, right of=input] (sum) {};
            %\node (in_text) at(3, 0.5) {Variable nodes};
            \node [var_node] at(0,0) (v1) {};
            \node[above of = v1, node distance = 0.33 cm] (in_text) at(3,0.5) (l1) {$Proc(3)$:\textit{MPI\_Recv()}};
            \node [var_node] at(1,0) (v2) {};
            \node [var_node] at(2,0) (v3) {};
            \node [var_node] at(3,0) (v4) {};
            \node [var_node] at(4,0) (v5) {};
            \node [var_node] at(5,0) (v6) {};
            \node [var_node] at(6,0) (v7) {};
            \node [var_node] at(7,0) (v8) {};
            \node [check_node] at(2,-1.5) (c1) {};
            \node [check_node] at(3,-1.5) (c2) {};
            \node [check_node] at(4,-1.5) (c3) {};
            \node [check_node] at(5,-1.5) (c4) {};
            \node[below of = c2, node distance = 0.33 cm] (in_text) at(3,-2) (l2) {$Proc(1)$:\textit{MPI\_Send()}};
            \node at(3, -3) (sub_cap) {Check node partial LLR ($E_{cv}$) update and data transfer};
            %\node (in_text) at(3, -2) {Check nodes};
            % We draw an edge between the controller and system block to 
            % calculate the coordinate u. We need it to place the measurement block. 
            %\node [output, right of=encoder] (output) {};
           % \node [right of=output] (out_text) at (3,0.25) {$p_1, p_2, p_3, ..., p_n$};
        
            % Once the nodes are placed, connecting them is easy. 
            %\draw [draw,->] (input) -- node {} (encoder);
            %\draw [->] (encoder) -- node {} (output);
            %\draw[->] (v1) -- node {} (c1);
            %\draw[->] (v3) -- node {} (c1);
            %\draw[->] (v5) -- node {} (c1);
            \draw[->] (c2) -- node {$E_{13}$} (v4);
            \draw[->] (v4) -- node {} (l1);
            \draw[->] (l2) -- node {} (c2);
        \end{tikzpicture}
    \end{center}
    
    \begin{tabular}{|c|c|}
        \hline
        
        \begin{tikzpicture}
            \node [var_node] at(0,0) {};
        \end{tikzpicture}
        &  Variable node \\
        \hline
         \begin{tikzpicture}
            \node [check_node] at(0,0) {};
        \end{tikzpicture}
        & Check node \\
        \hline
    \end{tabular}
    
    \caption{Example of communication between 4 processors while performing Sum-Product Decoding for Parity Check matrix shown in Fig. \ref{fig:graph}.}
    \label{fig:decoding}
\end{figure}

For Sum-Product Decoding, we create a graph and an adjacency list of the $\mathrm{H}$ matrix as shown in Section \ref{subsec:ldpc_decode}. Now, for each iteration of the decoding algorithm, we first divide the variable nodes and check nodes among all processors in an interleaved manner. For each iteration the processors calculate the partial LLR values $E_{cv}$ of check nodes for which the check node index $c\%N_{proc} = x$ using Eq. \ref{eq:extrin}. Then, after all processors have calculated their respective updated values, each processor sends the updated value of check nodes only to the processor with variable node index $v\%N_{proc} = x$. In the same way, the variable node LLR values are updated and sent. 
%The data transfer between processes takes place using the \textit{MPI\_Send()} and \textit{MPI\_Recv()} commands.
An example of communication between processors to update LLR values is shown in Fig. \ref{fig:decoding}. The decoding algorithm runs for preset number of iterations after which the variable nodes update their final LLR values and hard decision decoding is done using Eq. \ref{eq:hard_decision}. The complete Distributed Sum-Product Decoding is shown in Algorithm \ref{alg:decoding}. The LLR updation steps and the LLR transfer steps are separate for all processors. This separation ensures minimum switching between computation and communication for all processors.

\subsection{Distributed Batch Processing implementation}\label{subsec:batch_proc}

Batch Processing here means each processor takes a batch of vectors for encoding or decoding, and the number of total vectors of information bits is greater than the number of processors.

\subsubsection{{Encoding}}

Now, for encoding, we assume an input vector $M$ which is a $kN_e$ length input vector consisting of $k$ length vectors. Here $N_e$ is the number of $k$ length vectors to be encoded. The $M$ vector divided into smaller vectors of size $k$ which are then distributed among all processors. Each processor then uses Eq. \ref{eq:encoding} to encode the batch of $k$ length $\mathrm{m}$ vectors. The output of the batch of vectors at each processor is then gathered in the master processor using \textit{MPI\_Gatherv()}. The distribution of vectors is done using Eq. \ref{eq:elems_div}, where $l = N_e$.

\subsubsection{{Decoding}}

For decoding, a noisy vector of length $nN_d$ is taken, where $N_d$ is the number of vectors to be decoded. This vector is then divided into multiple $n$ length vectors which are distributed to multiple processors using Eq. \ref{eq:elems_div}. Then, each processor performs Sum-Product decoding, as shown in Section \ref{subsec:ldpc_decode}, on the batch of $n$ length vectors in a serial manner. The decoded output is then gathered in the master processor using \textit{MPI\_Gatherv()}.

%% file: Performance_Evaluation.tex
\section{Experiments and Results}\label{sec:eval}

\begin{table}[t]
    \centering
    \begin{tabular}{c|c}
        \hline
        \textbf{Parameter} & \textbf{Value} \\
        \hline
        Code length & 1032 \\
        Design Rate & $(0.25,0.5,0.75)$ \\
        Number of $1$s per row & 12 \\
        Number of $1$s per column & Depending on rate $(9, 6, 3)$ \\
        Number of decoding iterations & 10 \\
        Number of vectors processed & 1000 \\
        Number of servers & $1,2$ \\
        Number of CPUs per server & $2$ \\
        Number of processors per CPU & 1 to 32 \\
       % Modulation Scheme & BPSK \\
        \hline
    \end{tabular}
    \caption{Parameters for the experiments}
    \label{tab:exp_params}
\end{table}

\begin{figure}[t]
    \centering
    \begin{subfigure}{\linewidth}
    \includegraphics[width=\linewidth]{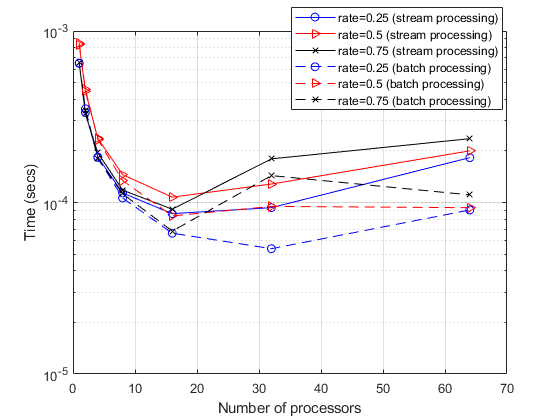}
    \subcaption{Encoding}
    \end{subfigure}
    \begin{subfigure}{\linewidth}
    \includegraphics[width=\linewidth]{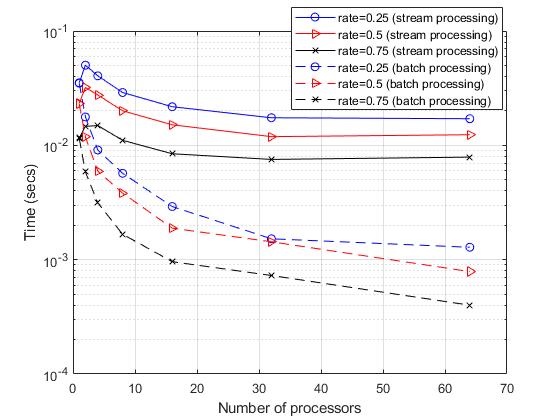}
    \subcaption{Decoding}
    \end{subfigure}
    \caption{Per-vector execution time for Stream Processing and Batch Processing of 1032 length LDPC codes with rates $(1/4, 1/2, 3/4)$ over single server}
    \label{fig:enc_dec_1_server_rate}
\end{figure}

\begin{figure}[t]
    \centering
    \begin{subfigure}{\linewidth}
    \includegraphics[width=\linewidth]{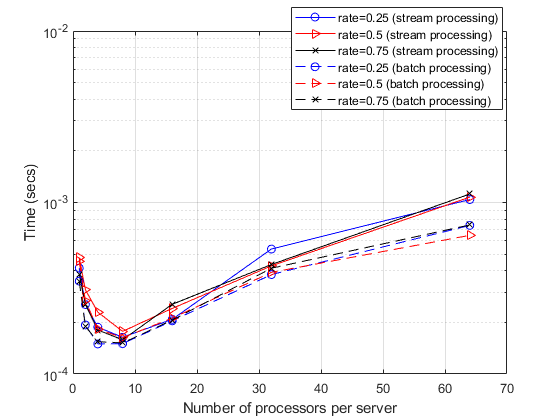}
    \subcaption{Encoding}
    \end{subfigure}
    \begin{subfigure}{\linewidth}
    \includegraphics[width=\linewidth]{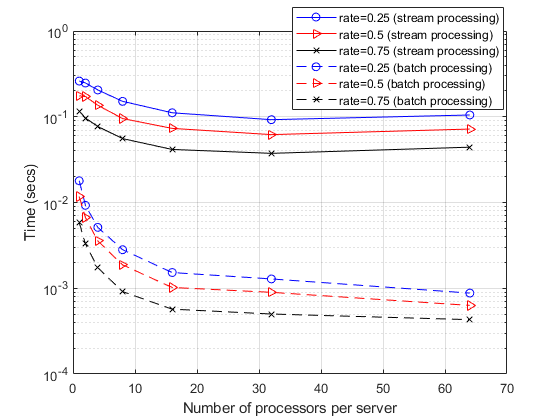}
    \subcaption{Decoding}
    \end{subfigure}
    \caption{Per-vector execution time for Stream Processing and Batch Processing of 1032 length LDPC codes with rates $(1/4, 1/2, 3/4)$ over two servers}
    \label{fig:enc_dec_2_server_rate}
\end{figure}

The parameters used for performing experiments are shown in Table \ref{tab:exp_params}. For the experiments, we design the Parity Check matrix for LDPC codes as shown in \cite{gallager_ldpc}. We design $ (N_c, 12)$ Parity Check matrix where $N_c$ is dependent on the code rate. Based on the designed matrix, encoding and decoding is performed as shown in previous sections. We conduct experiments in two parts. Firstly, we distribute the processing on a single server consisting of two CPUs. Secondly, we distribute the processing among all processors of two such servers. For both cases, we compare Stream Processing and Batch Processing based on the execution time taken for processing of a single input vector. To get per-vector execution time, the total execution time is divided by the number of vectors. %We run the experiments for different LDPC matrices of the same size (i.e code length) and then average the execution time.

\subsection{Single server case}

The execution time for distributed processing on multiple processors of a single server consisting of two CPUs is shown in Fig. \ref{fig:enc_dec_1_server_rate}. The difference in execution time between Stream and Batch Processing starts increases as the number of processors increase. Batch Processing starts performing better than Stream Processing because the processors only communicate for taking the input vector and giving the output vector. While for Stream Processing, the processors communicate after encoding each vector, and during each iteration of decoding of each vector. Also, when the number of processors is increased beyond the maximum logical cores per CPU, the performance of Stream Processing either saturates or worsens due to the added latency of inter-CPU communication.

Even though Batch Processing has lesser execution time per vector, if high efficiency in utilization of hardware resources is to be maintained, the number of vectors to be processed must always be greater than the number of processors. So, the initial latency of Batch Processing can be higher than Stream Processing when higher utilization efficiency is to be maintained.

\subsection{Two server case}

The execution time for distributed processing on multiple processors of two servers, with two CPUs each, is shown in Fig. \ref{fig:enc_dec_2_server_rate}. For two servers case, the x-axis of Fig. \ref{fig:enc_dec_2_server_rate} shows the number of processors used per server. Which means that if the x-axis shows $4$ then $4$ processors are being used per server i.e. total of $8$ processors are being utilized. It can be seen that Batch Processing performs better than Stream Processing, especially for decoding, due to the minimal communication latency between processors Also, the performance of Stream Processing using two servers for decoding is approximately an order of magnitude worse than performance of Stream Processing using a single server, and for encoding the performance worsens with increase in number of processors. So, for multi-server scenario, due to the high communication latency between servers, Stream Processing performance is dependent on the speed of inter-server communication links.

Since the communication between processors in Batch Processing is much lesser than the computation per processor, its dependence on the type of inter-server communication link is lesser as compared to that of Stream Processing, giving higher acceleration. Performance for decoding using Distributed Processing for two server case is either similar to or better than that of single server case with increase in number of processors. Performance of encoding using Distributed Processing for two server case is still worse than that of single server case. This is due to the computation time being negligible as compared to communication time.

%% file: Conclusion.tex
\section{Conclusion and future Work}\label{sec:conc}

Considering the intensive computation required for processing of LDPC codes, we used MPI to distribute the processing over LDPC codes over multiple multi-core CPUs. Using the distributed implementation, acceleration was provided for encoding and decoding processes of LDPC codes. Evaluation and comparison for Stream processing and Batch Processing based mechanisms for distributed processing of LDPC codes was done, and the advantages and limitations of both methods were shown.

While using a distributed memory based message passing model for distributed processing provides acceleration, increasing the number of processors to more than the logical cores in the system adds high processing latency as well as inter-processor communication latency due to context switching between processes. Also, for multi-server systems, the processing latency is highly dependent on the type of communication link between servers. So, to decrease the dependence on communication between CPUs, combination of shared memory and distributed memory based systems will be considered for distributed processing of LDPC codes.